\documentclass[prb,twocolumn,,floatfix]{revtex4}
\usepackage{graphicx}
\usepackage{bm}
\usepackage{amssymb}
\usepackage{color}

\newcommand{\be}{\begin{equation}}
\newcommand{\ee}{\end{equation}}
\def\prb{Phys. Rev. B }

\def \be{\begin{equation}}
\def \ee{\end{equation}}
\def \ba{\begin{array}}
\def \ea{\end{array}}
\def \bea{\begin{eqnarray}}
\def \eea{\end{eqnarray}}

\begin{document}

\title{A route to high temperature superconductivity in composite systems}
\author{Erez Berg$^1$, Dror Orgad$^2$ and Steven A. Kivelson$^1$}
\affiliation{$^1$ Department of Physics, Stanford University, Stanford, CA 94305-4045, USA%
\\
$^2$ Racah Institute of Physics, The Hebrew University, Jerusalem 91904,
Israel}

\begin{abstract}
Apparently, some form of local superconducting pairing persists to
temperatures well above the maximum observed $T_c$ in underdoped cuprates,
\textit{i.e.} $T_c$ is
suppressed 
due to the small phase stiffness. With this in mind, we consider the
following question -- 
Given a system with a high pairing scale $\Delta_0 $ but 
with $T_c$ reduced by phase fluctuations, 
can one design a composite system 
in which $T_c$ 
approaches its mean-field value, $T_c\to T_{MF}\approx \Delta_0/2\,$? Here,
we study a simple two component model 
in which a ``metallic layer" with $\Delta_0=0$ is coupled by single-particle
tunneling to a ``pairing layer" with 
$\Delta_0 >0 $ but zero phase stiffness. 
We show that in the limit that the bandwidth of the metal is much larger
than $\Delta_0$, $T_c$ of the composite system can reach the upper limit
$T_c \approx\Delta_0/2$.
\end{abstract}

\date{\today}
\maketitle

\section{Introduction}

There are both theoretical\cite{EmeryKivelson,ourreview} and experimental%
\cite{ourreview,Uemura,Corson,Ong} indications that underdoped cuprate
superconductors can exhibit significant pairing correlations for a range of
temperatures that extends above the highest measured superconducting $T_c$.
Whereas in conventional metallic superconductors, $T_c\approx \Delta_0/2$ is
determined by the pairing (zero $T$ gap) scale, in underdoped cuprates it is
apparently determined by the collective onset of phase coherence, and hence
by the superfluid stiffness, $\kappa \propto \rho_s$, where $\rho_s$ is the
zero $T$ superfluid density\cite{Uemura}.

The question we address here is: Given a material which has a
``high'' pairing scale, $\Delta_0$, but which fails to become a
superconductor at high temperatures due to its low superfluid
density, can we design an artificial composite of this material
and a simple metal that realizes a high transition temperature,
$T_c \sim \Delta_0/2\,$? Certainly superconductivity can be
induced in the simple metal via the proximity
effect\cite{merchant}, leading to an enhancement of the total
superfluid density. Conversely, however, the pairing scale tends
to be suppressed by the very same proximity
effect\cite{merchant,valles}. It is not clear, a priori, whether
the composite will exhibit the best of both worlds, or the worst.

Two sets of experimental observations suggest a positive outcome.
Firstly, in the last several years, Ong and
collaborators\cite{Ong} have shown that phenomena related to
fluctuation diamagnetism persist to moderately high temperatures
in underdoped cuprates. This, added to the older evidence that
there exists a spectroscopic pseudo-gap which extends to high
temperatures, encourages us to interpret at least a portion of the
observed ``pseudo-gap regime'' as a regime of pairing without
global phase coherence. Secondly, recent experiments by Yuli
\emph{et al.} \cite{Millo} on epitaxial films of
La$_{2-x}$Sr$_{x}$CuO$_{4}$ on a SrTiO substrate demonstrated that
$T_c$ of underdoped films may be raised by depositing a thin
upper-layer of strongly overdoped and hence metallic
La$_{1.65}$Sr$_{0.35}$CuO$_{4}$ (see also Ref.
[\onlinecite{bozovic}]).

Motivated by these findings, we study 
simple model systems composed of two components: a ``pairing" component with
a high pairing scale, $\Delta_0$, but zero $T_{c}$ due to zero superfluid
stiffness, and a ``metallic" component with no pairing but high stiffness.
The microscopic origin of the pairing is not elucidated in this work, and we
treat it as given. However, on physical grounds, we only consider situations
in which $\Delta_0 \ll E_F$, the Fermi energy of the metal. The two systems
are coupled by a tunnelling matrix element $t_{\perp } $. Our principle
result is the demonstration that, under the right conditions (\textit{i.e.}
the optimal magnitude of $t_{\perp }$), $T_{c}\approx \Delta_0/2$ can be
achieved. It is our hope that these results can provide guidance for a new
generation of searches, of the sort pioneered by Yuli \textit{et al.}\cite%
{Millo}, for higher temperature superconductivity in engineered composite
materials. More generally, this work extends previous work\cite%
{sudip,ekz,physica,arrigoni,martin,refael,hongwei,fradkin} on
``optimal inhomogeneity for superconductivity'' to situations more
amenable to direct experimental manipulation.


This paper is organized as follows. In Sec. \ref{sec:model} we describe the
model and our strategy of solving it. The results for the cases in which the
pairing layer consists of negative-$U$ sites and negative-$U$ wires are
presented in Sections \ref{sec:negU} and \ref{sec:scwires}, respectively.
The results are discussed in Sec. \ref{sec:discussion}.

\section{model and strategy}

\label{sec:model}

The ``pairing component" is modelled by a two dimensional lattice of
negative $U$ sites, which are either decoupled completely or coupled only in
one direction (forming an array of parallel one-dimensional wires). In both
cases, $T_{c}$ of the isolated pairing layer is zero due to zero phase
stiffness. Nevertheless, the system has a finite pairing scale $\Delta_0$.
Upon coupling this layer to a metallic layer modelled by non-interacting
electrons, a finite $T_{c}$ obtains. The behavior of $T_{c}$ as a function
of the strength of the coupling between the two systems is then studied.

The model Hamiltonian 
is
\begin{equation}
H=H_{c}+H_{f}+H_{cf}\text{,}  \label{H}
\end{equation}%
where $H_{c}$ is the Hamiltonian of the non-interacting (metallic) layer :
\begin{equation}
H_{c}=-t\sum_{\left\langle \mathbf{rr}^{\prime }\right\rangle \sigma }c_{%
\mathbf{r}\sigma }^{\dagger }c_{\mathbf{r}^{\prime }\sigma }^{%
\vphantom{\dagger}}+\mathrm{H.c.}-\mu \sum_{\mathbf{r}}n_{c,\mathbf{r}}\text{%
,}  \label{Hc}
\end{equation}%
where $\left\langle \mathbf{rr}^{\prime }\right\rangle $ denotes nearest
neighbors. $H_{f}$ is the Hamiltonian of the 
``pairing'' layer:
\begin{eqnarray}
H_{f} &=& -(\mu-\varepsilon) \sum_{\mathbf{r}}n_{f,\mathbf{r}}  \nonumber \\
&&-U\sum_{\mathbf{r}}\left( n_{f,\mathbf{r}\uparrow }-\frac{1}{2}\right)
\left( n_{f,\mathbf{r}\downarrow }-\frac{1}{2}\right)  \nonumber \\
&&-t^{\prime }\sum_{\mathbf{r\sigma }}f_{\mathbf{r}\sigma }^{\dagger }f_{%
\mathbf{r}+a\mathbf{\hat{x}},\sigma }^{\vphantom{\dagger}}+\mathrm{H.c.}
\label{Hf}
\end{eqnarray}%
with $U>0$ (attractive) and $t^\prime=0$ for the ``pairing sites'' problem
analyzed in Sec. \ref{sec:negU}, and $t^\prime=t $ 
for the superconducting wires problem analyzed in Sec. \ref{sec:scwires}.
Finally, $H_{cf} $ is the the tunneling Hamiltonian between the two layers,
\begin{equation}
H_{cf}=-t_{\perp }\sum_{\mathbf{r}\sigma }c_{\mathbf{r}\sigma }^{\dagger }f_{%
\mathbf{r}\sigma }+\mathrm{H.c.}  \label{Hfc}
\end{equation}%
Here $c_{\mathbf{r}\sigma }^{\dagger }$ and $f_{\mathbf{r}\sigma }^{\dagger
} $ create electrons in the
metallic and pairing layers, respectively, $n_{\mathbf{r}}=$ $n_{f,\mathbf{r}%
}+n_{c,\mathbf{r}}$ where $n_{f,\mathbf{r}}=\sum_{\sigma =\uparrow
,\downarrow }$ $f_{\mathbf{r}\sigma }^{\dagger }$ $f_{\mathbf{r}\sigma }$,
and similarly for $n_{c,\mathbf{r}}$. The on-site energy $\varepsilon $ for
the $f$ sites is assumed to be close to the chemical potential, so that the $%
f$ band is partially filled. Throughout the analysis, we assume that the
metallic bandwidth $W=8t$ is much larger than 
$t_{\perp }$, $U$. 
For simplicity, we consider the case of a two-dimensional square lattice,
with a lattice constant $a=1$.

In order to solve (\ref{H}), we first use mean-field theory to decouple the
interaction term:
\begin{equation}
-Uf_{\mathbf{r\downarrow }}^{\dagger }f_{\mathbf{r\uparrow }}^{\dagger }f_{%
\mathbf{r\uparrow }}^{\vphantom{\dagger}}f_{\mathbf{r\downarrow }}^{%
\vphantom{\dagger}}\rightarrow -\Delta ^{\ast }f_{\mathbf{r\uparrow }}^{%
\vphantom{\dagger}}f_{\mathbf{r\downarrow }}^{\vphantom{\dagger}}-\Delta f_{%
\mathbf{r\downarrow }}^{\dagger }f_{\mathbf{r\uparrow }}^{\dagger }+\delta
\varepsilon n_{f,\mathbf{r}}\text{,}
\end{equation}%
and solve the self-consistent BCS\ equations at finite temperature:
\begin{equation}
\Delta =U\left\langle f_{\mathbf{r\uparrow }}f_{\mathbf{r\downarrow }%
}\right\rangle ,  \label{bcs1}
\end{equation}%
\begin{equation}
\delta \varepsilon =-\frac{U}{2}\langle n_{f,\mathbf{r}}\rangle ,
\end{equation}%
\begin{equation}
n=\left\langle n_{\mathbf{r}}\right\rangle \text{,}  \label{bcs2}
\end{equation}%
where $n$ is some fixed density. From these equations we find the mean-field
transition temperature $T_{MF}$, at which $\Delta $ vanishes. However, the
actual $T_{c}$ of the model 
is lower than $T_{MF}$ due to phase fluctuations, which are particularly
important in situations where the phase stiffness is small, \emph{i.e.} when
$t_{\perp }$ is small. (Note that when $t_{\perp }=0$, $T_{MF}>0$, but $%
T_{c}=0$. This is true regardless of $t^{\prime }$.)

We make an estimate of the superconducting $T_{c}$ that includes both the
usual physics of pairing that is captured by BCS mean-field theory and the
dominant effects of phase fluctuations, as follows:
To begin with, we compute the mean-field approximation to the phase
stiffness $\rho _{s}\left( T\right) $, defined as
\begin{equation}
\rho _{s}\left( T\right) =\frac{1}{\Omega }\frac{\partial ^{2}F}{\partial
q_{x}^{2}}\text{,}  \label{rs}
\end{equation}%
where $F/\Omega $ is the free energy per unit area and $q_{x}$ is a phase
twist in the $x$ direction, 
which enters the kinetic energy term in the Hamiltonian as:%
\begin{equation}
-t\sum_{\left\langle \mathbf{rr}^{\prime }\right\rangle \sigma }c_{\mathbf{r}%
\sigma }^{\dagger }c_{\mathbf{r}^{\prime }\sigma }^{\vphantom{\dagger}%
}\rightarrow -t\sum_{\left\langle \mathbf{rr}^{\prime }\right\rangle \sigma
}e^{i\frac{\mathbf{q}}{2}\mathbf{\cdot }\left( \mathbf{r}^{\prime }-\mathbf{r%
}\right) }c_{\mathbf{r}\sigma }^{\dagger }c_{\mathbf{r}^{\prime }\sigma }^{%
\vphantom{\dagger}}
\end{equation}%
(Eq. (\ref{rs}) is slightly modified in cases where $t^{\prime }\neq 0$,
since then the stiffness is anisotropic, and the relevant quantity is the
geometric mean of the stiffness in the $x$ and $y$ directions. This will be
discussed in Sec. \ref{sec:scwires}.)
Then, we estimate the temperature at which the two dimensional
Kosterlitz-Thouless transition (phase ordering) occurs in terms of 
the universal jump of the stiffness at criticality: 
\begin{equation}
\rho _{s}\left( T_{c}\right) =\frac{2}{\pi }T_{c}.  \label{KT}
\end{equation}%
This is still an overestimate as it 
neglects the renormalization of $\rho _{s}\left( T_{c}\right) $ due to phase
fluctuations below $T_{c}$. Upon solving Eqs. (\ref{bcs1}-\ref{bcs2},\ref{KT}%
), we estimate $T_{c}$ as a function of the model parameters. 
Although $T_{c}$ estimated in this way is always less than $T_{MF}$, where $%
\rho _{s}\left( T\right) $ vanishes, if the phase stiffness is very large
(as in a conventional weakly coupled BCS superconductor), then $T_{c}\approx
T_{MF}$.

The method described above to determine $T_c$
was applied in Ref. [\onlinecite{denteneer}] for the negative $U$ Hubbard
model, and the results 
were compared with the results of 
Quantum Monte Carlo (QMC) simulations
\cite{moreo_scalapino,scalettar}. Qualitative trends of the Monte-Carlo
results at generic fillings were well reproduced by this method\cite%
{note-half-filling}. Moreover, although the Monte-Carlo $T_c$ was always
smaller than the estimated $T_c$, the two typically differ by no more than
30\% - 50\%. Therefore, even though the method is not quantitatively
reliable in the intermediate to strong coupling regime, we do expect it to
predict correctly the qualitative trends of $T_c$ as a function of the model
parameters. We intend to check the results using Monte-Carlo methods in the
future.

\section{negative U sites}

\label{sec:negU}

Let us focus on the case $t^{\prime }=0$ in Eq. (\ref{Hf}), in which the
negative-$U$ sites are coupled only by tunnelling through the metallic
layer. We fix $t$, $U$, $\varepsilon $ and $n$, always assuming that
$U$, $t_{\perp }\ll W$, where $W=8t$ is the bandwidth of the
metallic layer, and calculate $T_{c}\left( t_{\perp }\right) $ [$\mu $ is
determined by Eq. (\ref{bcs2})]. $n$ is chosen so that the band of negative $%
U$ sites is partially filled (so that the self-consistent solution satisfies
$\mu \approx \varepsilon +\delta \varepsilon $ and $\Delta \neq 0$ at $T=0$).

\subsection{Analytical results}

In the limit $t_{\perp }$, $T\ll U$, the dependence of $T_{c}$ on $t_{\perp
} $ can be understood analytically from (4$^{th}$ order) perturbation theory
in $t_\perp$. In this limit, we may assume that $\Delta $ is approximately
temperature independent and equal to its zero temperature value $\Delta
_{0}\approx U/2$. 
At $T=0$, the perturbative expression is complicated due to Fermi surface
singularities, but for temperatures in the important range $\Delta_0 \gg
T\gg t_{\perp }^{2}/\Delta _{0}$, 
the results simplify [see Appendix A, Eq. (\ref{rs_low_app})]:
\begin{equation}
\rho _{s}\left( \frac{t_{\perp }^{2}}{U}\ll T\ll U\right) \sim \frac{%
t_{\perp }^{4}}{U^{2}T^{2}}\left\langle \mathbf{v}_{F}^{2}\right\rangle
_{FS}N\left( 0\right) ,  \label{rs_low_tperp}
\end{equation}%
where $N\left( 0\right) $ is the density of states of the metallic layer at
the Fermi energy, and $\left\langle \mathbf{v}_{F}^{2}\right\rangle _{FS}$
is the square of its Fermi velocity averaged over the Fermi surface.
Numerical factors of the order of unity have been dropped.
Since parametrically $\left\langle \mathbf{v}_{F}^{2}\right\rangle
_{FS}N\left( 0\right) \sim t$, this gives $\rho _{s}\left( T\right) \sim
\frac{t_{\perp }^{4}}{U^{2}T^{2}}t$. Using Eq. (\ref{KT}), we get the
following estimate of $T_{c}$:
\begin{equation}
T_{c}\left( t_{\perp }\ll U\right) \sim t\left( \frac{t_{\perp }}{\sqrt{Ut}}%
\right) ^{\frac{4}{3}}\text{.}  \label{tc_low}
\end{equation}%
Eq. (\ref{tc_low}) gives that $T_{c}\gg $ $t_{\perp }^{2}/U$, consistent
with the assumptions leading to Eq. (\ref{rs_low_tperp}). Eq. (\ref{tc_low})
must break down before $t_{\perp }\simeq t_{\perp ,1}$ where
\begin{equation}
t_{\perp ,1}\sim U\left( \frac{U}{t}\right) ^{\frac{1}{4}}  \label{tp1}
\end{equation}%
since Eq. (\ref{tc_low}) gives $T_{c}\left( t_{\perp ,1}\right) \sim U\sim
\Delta _{0}$, and $T_{c}$ cannot exceed $\Delta _{0}$.

As $t_{\perp }$ is increased beyond $t_{\perp ,1}$, the superfluid density
is large enough and ceases to limit $T_{c}$ significantly. However, the
pairing is also reduced. $T_{MF}$, the temperature at which $\Delta \left(
T_{MF}\right) =0$, can be calculated perturbatively in $t_{\perp }$
[Eq. (\ref{TMF_low_app}) of Appendix A]:
\begin{equation}
T_{MF}=\frac{U}{4}\left[ 1-\frac{At_{\perp }^{2}}{Ut}+O\left( t_{\perp
}^{4}\right) \right],  \label{TMF_low}
\end{equation}%
where, to be explicit, 
we have taken the negative $U$ sites to be half filled for $t_{\perp }=0$. $%
A $ is a dimensionless number of order unity. Therefore $T_{MF}$ is not
suppressed significantly from its $t_{\perp }=0$ limit until $t_{\perp }$
becomes of the order of
\begin{equation}
t_{\perp ,2}\sim \sqrt{Ut}.  \label{tp2}
\end{equation}

Interestingly, we see that in the limit $U\ll t$, the ratio $\frac{t_{\perp
,2}}{t_{\perp ,1}}=\left( \frac{t}{U}\right) ^{\frac{3}{4}}$ becomes large.
Therefore, there is a \emph{parametrically wide} region where there is
plenty of superfluid stiffness, but the pairing is still not suppressed
significantly. It is at least plausible to expect that in the region $%
t_{\perp ,1}\lesssim t_{\perp }\lesssim t_{\perp ,2}$, $T_{c}$ of the order
of $T_{MF}\left( t_{\perp }=0\right) \approx 
{\Delta _{0}}/{2}$ is obtained.

\subsection{Numerical results}


\begin{figure}[t]
\centering
\includegraphics[width=8.6cm]{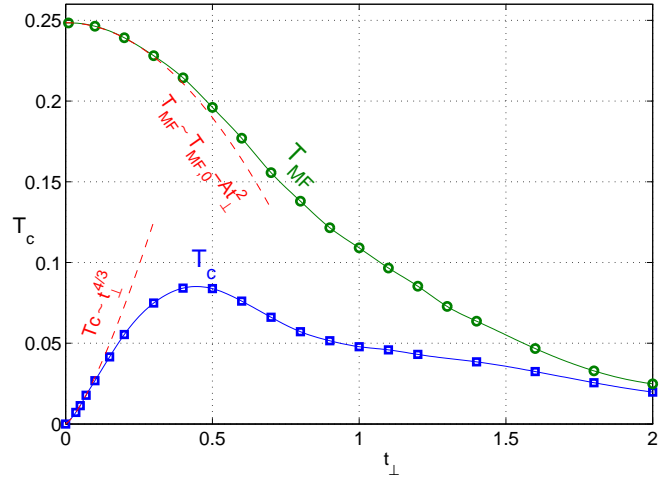}
\caption{(Color online.) $T_c$ ($\square$) and $T_{MF}$ ($\circ$) obtained
from solving Eqs. (\protect\ref{bcs1}-\protect\ref{bcs2},\protect\ref{rs})
numerically for $n=1.5$, $t=1$, $U=1$, as a function of $t_\perp$. The
dashed curves are fits to the data according to Eqs. (\protect\ref{tc_low},%
\protect\ref{TMF_low}). $A=0.235$ was used in the fit.}
\label{fig:Tc_tperp_Ueq1}
\end{figure}
\begin{figure}[ht]
\centering \vspace{-0cm} %
\includegraphics[width=8.6cm]{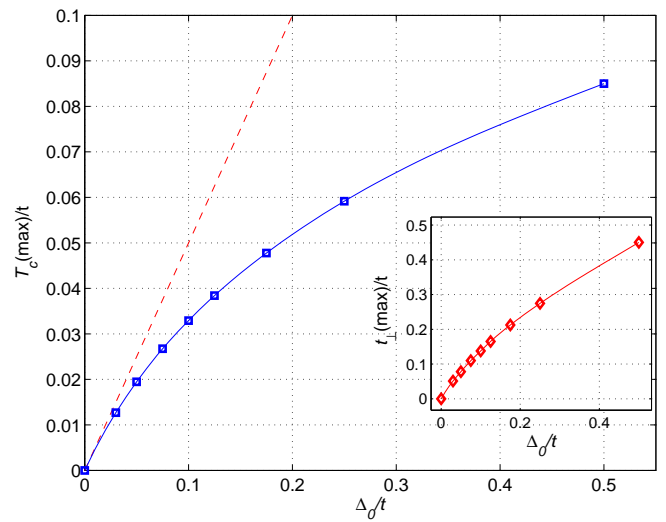}
\caption{(Color online.) $T_c$(max) (maximized over $t_\perp$) for different
values of $U$ and fixed $t=1$, $n=1.5$. $T_c$(max) is shown as a function of
$\Delta_0\approx U/2$, which is the $T=0$, $t_\perp=0$ gap for the same $U$.
The dashed line is the mean field transition temperature for $t_\perp=0$, $%
T_{MF,0}\approx\Delta_0/2$. \emph{inset:} $t_\perp$(max) in which $T_c$(max)
is obtained, as a function of $\Delta_0$.}
\label{fig:Tc_max}
\end{figure}

Fig. \ref{fig:Tc_tperp_Ueq1} shows $T_c$ and $T_{MF}$ obtained from solving
Eqs. (\ref{bcs1}-\ref{bcs2},\ref{rs}) numerically for $n=1.5$, $t=1$, $%
\varepsilon=-1$ and $U=1$, as a function of $t_\perp$. At low $t_\perp$, $%
T_{MF}\approx U/4$, while $T_c $ is strongly suppressed due to the low
superfluid stiffness. For low enough $t_\perp$, $T_c \sim t_\perp^{4/3}$ in
agreement with Eq. (\ref{tc_low}). $T_c $ reaches a maximum at $t_\perp
\approx 0.45$, and then starts to drop due to the suppression of $T_{MF}$.
At high enough $t_\perp$, $T_c$ essentially coincides with $T_{MF}$. The
maximum $T_c$, which is obtained in the crossover regime between
pairing-dominated and stiffness-dominated regimes, is $T_c \approx 0.085$,
which is about $35\%$ of the maximum $T_{MF}$. 

In Fig. \ref{fig:Tc_max} we show $T_{c}\left( \max \right) $, which is the
maximum of $T_{c}\left( t_{\perp }\right) $, as a function of $\Delta
_{0}\approx U/2$, which is the $T=0$, $t_{\perp }=0$ gap. We fix $t=1$ and $%
n=1.5$ throughout the calculation. In the low $\Delta _{0}/t$ limit, $%
T_{c}\left( \max \right) $ reaches the maximum conceivable value which takes
full advantage of the pairing scale, 
$T_{c}\left( \max \right)/ \Delta _{0}\to A_0\approx 1/2$ as $\Delta_0/t\to
0 $. (The dashed line in Fig. \ref{fig:Tc_max} is $T_{c}\left( \max \right)=
\Delta _{0}/2$).

The optimal $t_{\perp }$ for superconductivity, $t_{\perp }$(max) is shown
in the inset of Fig. \ref{fig:Tc_max} as a function of $\Delta _{0}$. For
small $\Delta _{0}/t$, we find that $t_{\perp }$(max)$\approx \Delta _{0}$.
As $\Delta _{0}$ is lowered, the maximum becomes broader and broader
relative to $\Delta_0$, in agreement with what we expect from Eq. (\ref{tp1},%
\ref{tp2}): $T_{c}\approx \Delta_{0}/2$ for $t_{\perp ,1}\lesssim t_{\perp
}\lesssim t_{\perp ,2}$, and this range becomes parametrically wide at low $%
\Delta_{0}$.

\section{superconducting wires}

\label{sec:scwires}

\subsection{Analytical results}

\begin{figure}[b]
\centering
\includegraphics[width=8.6cm]{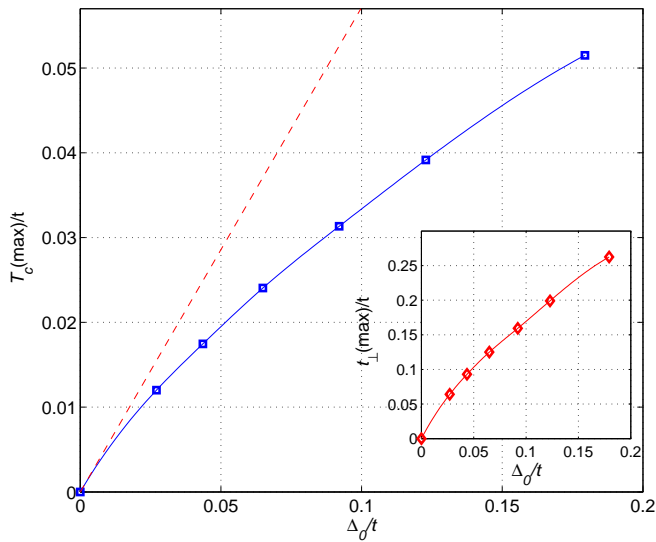}
\caption{(Color online.) Same as Fig \protect\ref{fig:Tc_max} for the case
of superconducting wires [$t^{\prime}=t$ in Eq. (\protect\ref{Hf})]. $T_c$%
(max)$/t$ (maximized over $t_\perp$) is shown as a function of $\Delta_0/t$.
The dashed line shows the mean-field transition temperature at $t_\perp=0$, $%
T_{MF,0}/t\approx2\Delta_0$/(3.5t). \emph{inset:} $t_\perp$(max)$/t$ as a
function of $\Delta_0/t$.}
\label{fig:Tc_max_1D}
\end{figure}

We now consider the case in which the \textquotedblleft pairing layer" is an
array of one dimensional wires in the $x$ direction.
Assuming that $t^{\prime }\sim t>U$, 
the zero temperature gap is given by the BCS\ equation: $\Delta _{0}\sim
t^{\prime }\exp [
-1/\tilde{N}(0)U]$ where $\tilde{N}(0)\sim (2\pi t^{\prime })^{-1}$ is the
density of states of a single wire. The phase stiffness along the $x$
direction is finite even for $t_{\perp }\rightarrow 0$,\cite{comment-wires1}
while the stiffness in the $y$ direction vanishes at this limit. Since the
phase stiffness is anisotropic, $\rho _{s}^{x}\neq \rho _{s}^{y}$, the
macroscopic phase stiffness, which appear in Eq. (\ref{KT}), is an
appropriate average of its values in the two directions.
Analogously to the case of the anisotropic two dimensional $XY$ model, the
geometric average of $\rho _{s}^{x}$, $\rho _{s}^{y}$ should be used\cite%
{dimcross}:
\begin{equation}
\rho _{s}=\sqrt{\rho_{s}^{x}\rho_{s}^{y}}\text{.}
\end{equation}

We have found that $T_c$ of the composite system is highest when the Fermi
surfaces of the two layers 
intersect in the $t_\perp \to 0$, $U \to 0$ limit \cite{comment-wires2}. We
therefore assume that this is the case in what follows.

Following a similar line of reasoning as in Sec. \ref{sec:negU}, the scaling
of $\rho _{s}$ in the limit $\frac{t_{\perp }^{2}}{\Delta _{0}}\ll T\ll
\Delta _{0}$ is [Eq. (\ref{rs_low_app_wires}) in Appendix \ref%
{sec:app_low_tperp}]:
\begin{equation}
\rho _{s}\left( \frac{t_{\perp }^{2}}{\Delta _{0}}\ll T\ll \Delta
_{0}\right) \sim \sqrt{\frac{t}{\Delta _{0}}}\frac{t_{\perp }^{2}}{T}\text{,}
\label{rs_low_wires}
\end{equation}%
which by Eq. (\ref{KT}) gives
\begin{equation}
T_{c}\left( t_{\perp }\ll \Delta _{0}\right) \sim \left( \frac{t}{\Delta _{0}%
}\right) ^{\frac{1}{4}}t_{\perp }\text{.}  \label{tc_low_wires}
\end{equation}%
And, since $T_{c}$ cannot exceed $\sim \Delta _{0}$, Eq. (\ref{tc_low_wires}%
) can only hold for
\begin{equation}
t_{\perp }\lesssim t_{\perp ,1}=\left( \frac{\Delta _{0}}{t}\right) ^{\frac{1%
}{4}}\Delta _{0}\text{.}
\end{equation}

The small $t_{\perp }$ behavior of $T_{MF}$ is [Eq. (\ref{TMF_app_low_1D})]
\begin{equation}
T_{MF}\approx T_{MF,0}\left[ 1-\frac{\tilde{A}t_{\perp }^{2}}{tT_{MF,0}}%
+O\left( t_{\perp }^{4}\right) \right] \text{,}
\end{equation}%
where $T_{MF,0}=T_{MF}\left( t_{\perp }=0\right) \approx {\Delta _{0}}/{2}$
and $\tilde{A}$ is a dimensionless constant of order unity. Therefore, the
suppression of $T_{MF}$ due to the coupling of the superconducting wires to
the metallic layer becomes significant when
\begin{equation}
t_{\perp }\gtrsim t_{\perp ,2}=\sqrt{\Delta _{0}t}\text{.}
\end{equation}%
Thus, as in the case of isolated negative-$U$ sites, there is a region
between $t_{\perp ,1}$ and $t_{\perp ,2}$ where there is plenty of phase
stiffness and the pairing is not suppressed significantly. 
Moreover, since $
{t_{\perp ,2}}/{t_{\perp ,1}}\sim \left( {t}/{\Delta _{0}}\right) ^{\frac{3}{%
4}}$, this region becomes parametrically wide when $\Delta _{0}\ll t$. In
that limit, we expect that $T_{c}$ can be asymptotically close to $T_{MF,0}$.

\subsection{Numerical results}

Fig. \ref{fig:Tc_max_1D} shows $T_{c}$(max) (maximized over $t_\perp$) as a
function of $\Delta_0$ for the case of 1D wires. The following parameters
were used: $t=t^{\prime}=1$, $\varepsilon=-1$ and $n=1.5$. $U$ was varied
between $1.1$ and $1.65$. Also shown in the same figure is $T_{MF,0}$, the
mean-field transition temperature of the wires for $t_\perp=0$. We found
that in the range of $U$ we considered, $T_{MF,0}$ is very well approximated
by the BCS formula $T_{MF,0}=2\Delta_0/3.5=a\exp(-b/U)$, with $a=4.415$ and $%
b=6.215$, \textit{i.e.} $\Delta_0$ changes by an order of magnitude from $%
\Delta_0\approx3\times 10^{-2}$ to $1.8\times 10^{-1}$. As in the case of
the negative $U$ sites, in the limit $\Delta_0/W \to 0$, $T_{c}$(max)
approaches $T_{MF,0}$. 

The inset of Fig. \ref{fig:Tc_max_1D} shows the optimal value of $t_\perp$
as a function of $\Delta_0$. For small $\Delta_0$, we see that $t_\perp$(max)%
$\approx 2\Delta_0$.



\section{discussion}

\label{sec:discussion}

The pairing scale $\Delta_0$ defines a 
physical limit on the maximum achievable superconducting $T_c$ 
in a given system. However, typically as the phase stiffness is increased,
the pairing scale tends to be suppressed,
and eventually this suppresses the actual $T_c$. Therefore, the maximum $T_c$
is typically reduced relative to $\Delta_0$, often by a large factor.
For example, in the two-dimensional negative-$U$ Hubbard model with fixed $U$%
, the maximum possible $\Delta_0$ is about $U/2$, which is achieved for $t=0$
and close to half filling. However, the maximum $T_c$ (estimated by the
method of combining the mean-field solution with classical phase
fluctuations, as described in Sec. \ref{sec:model}) is only $0.085U$
(obtained for $t\approx 0.4U$). 




In the present work, we have been motivated by the following question:
Suppose that there exists a material with a large pairing scale, $\Delta
_{0} $, but a low (or vanishing) $T_{c}$ due to phase fluctuations; is there
a way to make a composite of this material and a good metal which will
realize a superconducting state with a transition temperature, $T_{c}\to
T_{MF,0} \approx \Delta _{0}/2\,$? In the 
{two} model systems we studied, we found that by weakly coupling the two
materials with $t_{\perp }\sim \Delta _{0}$, and in the limit that the
bandwidth of the metal is large, $W/\Delta _{0}\rightarrow \infty $, this
optimal $T_{c}$ can be achieved.

This result was demonstrated using a physically motivated approximate
solution of the model. Fortunately, the negative $U$ Hubbard model is
amenable to solution on moderately large systems by Quantum Monte Carlo
Methods, as it can be made free of fermion sign problems\cite{dossantos}. We
therefore intend to test the validity of our results in this way in the near
future.

Finally, we discuss the reasons to believe that our conclusions do not
depend sensitively on the specifics of the models. The coupling of a paired
material to a good metal produces two qualitatively different effects:
an increased superfluid stiffness, $\delta \rho _{s}$, 
and a reduction of the mean field transition temperature by an amount $%
\delta T_{MF}$. It is clear that in the limit of strong coupling between the
two systems, $t_{\perp }\sim W$ where $W$ is the metallic bandwidth, the
latter effect always dominates, and hence coupling to the metal leads to a
\emph{quenching} of superconductivity.

Let us therefore consider $t_\perp \sim \Delta_0 \ll W$, where
a perturbative expression for $\delta T_{MF}$ 
will generally give 
\begin{equation}
\delta T_{MF}=-A W^{-a}\Delta_0^{a-1}t_\perp^2 \sim -W^{-a}\Delta_0^{a+1}.
\end{equation}
where 
$A>0$ is a dimensionless constant, $a$ is an exponent which could differ
from case to case, and in the final expression we have taken $t_\perp \sim
\Delta_0$. 
Similarly, 
close to the putative superconducting transition temperature $%
T\approx\Delta_0/2$, we expect
\begin{equation}
\delta \rho _{s}=B W^{b}\Delta_0^{-1-b}|\Delta(T)|^{2},  \label{rs_low_Delta}
\end{equation}
where $B$ is another constant, and $|\Delta(T)| \ll \Delta_0$ is the
temperature dependent mean field gap in the pairing layer. So long as $a>0$
and $b>0$, these relations imply that in the limit that $W\to\infty$, the
induced phase stiffness 
at any $T < T_{MF,0}$ 
grows without bound 
with no significant loss of pairing.
Hence 
phase fluctuations are suppressed, leading to $T_c \to T_{MF,0}$.

Generally, one expects that $\delta \rho_s$ increases as $W$ is increased (%
\textit{i.e.} $b>0$), while $|\delta T_{MF}|$ decreases ($a>0$), since, in
the metal, the Fermi velocity is a linearly increasing function and the
Fermi energy density of states is a linearly decreasing function of $W$.
Indeed, in the case of negative $U$ sites, $a= 
b=1$, \cite{comment-exponents} a result which, we believe, is true in a wide
range of circumstances.

As corroborating evidence, we note that the expected non-monotonic
dependence of $T_c$ on coupling between a metal and a phase fluctuating
superconductor has been observed in a somewhat analogous experimental system%
\cite{merchant} consisting of Pb grains covered with a film of Ag. As a
function of increasing Ag coverage, the first effect is to suppress phase
fluctuations and to increase the superconducting transition temperature up
to nearly the bulk $T_c$ of Pb. \cite{comment-merchant} However, adding more
Ag to the system eventually causes a degradation of the pairing scale and a
total quenching of superconductivity.

As a concluding remark, we comment on the effect of the pairing symmetry on
our results. So far we have considered cases where the superconducting order
parameter has $s$-wave symmetry. In the case of $d$-wave symmetry, the
induced order parameter in the metal has nodes. This will reduce the
superfluid density at low temperature relative to the $s$-wave case, due to
the excitation of nodal quasi-particles. However, at $T\approx \Delta_0/2$,
the behavior of $\rho_s$ is not expected to be qualitatively different from
the $s$-wave case. Therefore we expect our main results, $T_c$(max)$\approx
\Delta_0/2$, to hold in the $d$-wave case as well. We intend to test this
claim explicitly in the future.

\acknowledgments{We thank O. Millo, T. Pereg-Barnea, R. T.
Scalettar, W-F. Tsai, and H. Yao for their comments on this
manuscript. E. Altman and T. H. Geballe are acknowledged for many
stimulating discussions. This work was supported by the United
States - Israel Binational Science Foundation (grant No. 2004162),
by D.O.E. grant \# DE-FG02-06ER46287, and by the the Israel Science
Foundation (grant No. 538/08).}

\appendix

\section{The low $t_{\perp }$ limit}

\label{sec:app_low_tperp}

\subsection{Superfluid density}

We will now derive Eqs. (\ref{rs_low_tperp},\ref{rs_low_wires}) for the
superfluid density in the limit $\frac{t_{\perp }^{2}}{\Delta _{0}}\ll T\ll
\Delta _{0}$, where $\Delta _{0}$ is the zero temperature gap in the
\textquotedblleft pairing layer". In this limit, we assume that $\Delta
\left( T\right) =\Delta _{0}$ is independent of temperature. We proceed by
integrating out the (gapped) negative-$U$ layer degrees of freedom,
obtaining an effective action for the metallic layer. Focusing on the low
energy modes of the metallic layer, the $\omega $ dependence of the
effective action can be neglected, obtaining a low-energy effective
Hamiltonian of the form
\begin{equation}
H_{\text{eff}}=\sum_{\mathbf{k}\sigma }\xi _{\mathbf{k}+\frac{\mathbf{q}}{2}%
}c_{\mathbf{k}\sigma }^{\dagger }c_{\mathbf{k}\sigma }^{\vphantom{\dagger}}+%
\tilde{\Delta}\sum_{\mathbf{k}}c_{\mathbf{k}\uparrow }^{\dagger }c_{\mathbf{%
-k}\downarrow }^{\dagger }+\mathrm{H.c.}\text{,}  \label{Heff}
\end{equation}%
where $\xi _{\mathbf{k}}=-2t\left( \cos k_{x}+\cos k_{y}\right) -\mu $, $%
\mathbf{q}/2$ is a vector potential introduced in order to calculate the
phase stiffness, and%
\begin{equation}
\tilde{\Delta}\sim \frac{t_{\perp }^{2}}{\Delta _{0}}  \label{Dprox}
\end{equation}%
is the proximity induced pairing field in the metallic layer. Note that in
the $T\ll \Delta _{0}$ limit, $\tilde{\Delta}$ is approximately temperature
independent, even for temperatures larger than $\tilde{\Delta}$. The phase
stiffness at temperature $T$ is calculated from (\ref{Heff}) in the standard
way by computing the free energy $F\left( \mathbf{q}\right) =-T\ln Z\left(
\mathbf{q}\right) $ where $Z\left( \mathbf{q}\right) =$Tr$\left[ \exp \left(
-\beta H_{\text{eff}}\right) \right] $, and differentiating twice the free
energy per unit area with respect to $q_{x}$. This gives\cite%
{white_zhang_scalapino}
\begin{eqnarray}
\rho _{s} &=&\frac{1}{2\Omega }\sum_{\mathbf{k}}\left[ u_{\mathbf{k}%
}^{2}f\left( E_{\mathbf{k}}\right) +v_{\mathbf{k}}^{2}\left( 1-f\left( E_{%
\mathbf{k}}\right) \right) \right] \frac{\partial ^{2}\xi _{\mathbf{k}}}{%
\partial k_{x}^{2}}  \nonumber \\
&&-\frac{1}{2\Omega }\sum_{\mathbf{k}}\left( \frac{\partial \xi _{\mathbf{k}}%
}{\partial k_{x}}\right) ^{2}\beta f\left( E_{\mathbf{k}}\right) \left[
1-f\left( E_{\mathbf{k}}\right) \right] \text{,}  \label{rhos_BCS}
\end{eqnarray}%
where $E_{\mathbf{k}}=\sqrt{\tilde{\Delta}^{2}+\xi _{\mathbf{k}}^{2}}$, $u_{%
\mathbf{k}}=\sqrt{\frac{1}{2}\left( 1+\frac{\xi _{\mathbf{k}}}{E_{\mathbf{k}}%
}\right) }$, $v_{\mathbf{k}}=\sqrt{\frac{1}{2}\left( 1-\frac{\xi _{\mathbf{k}%
}}{E_{\mathbf{k}}}\right) }$ and $f\left( \varepsilon \right) $ is the Fermi
function. Integrating the first line of Eq. (\ref{rhos_BCS}) by parts and
replacing $\int \frac{d^{2}k}{\left( 2\pi \right) ^{2}}\rightarrow \int d\xi
N\left( \xi \right) $, where $N\left( \xi \right) $ is the density of states
of the metallic layer, we get
\begin{eqnarray}
\rho _{s} &=&\frac{1}{2}\int_{-W/2-\mu }^{W/2-\mu }d\xi N\left( \xi \right)
\frac{\tilde{\Delta}^{2}}{E^{2}}{\Bigg\{}\frac{1}{2E}\tanh \left( \frac{%
\beta E}{2}\right)  \nonumber \\
&&-\beta f\left( E\right) \left[ 1-f\left( E\right) \right] {\Bigg \}}%
\left\langle \mathrm{v}_{x}^{2}\right\rangle \left( \xi \right) \text{.}
\label{rhos_xi}
\end{eqnarray}%
Here $E\left( \xi \right) =\sqrt{\xi ^{2}+\tilde{\Delta}^{2}}$ and the
averaged square velocity at energy $\xi $ of the metallic layer is $%
\left\langle \mathrm{v}_{x}^{2}\right\rangle \left( \xi \right) =\frac{1}{%
N\left( \xi \right) }\int \frac{d^{2}k}{\left( 2\pi \right) ^{2}}\delta
\left( \xi _{\mathbf{k}}-\xi \right) \left( \frac{\partial \xi _{\mathbf{k}}%
}{\partial k_{x}}\right) ^{2}$. Assuming $|\pm W/2-\mu |\gg T$ implies that
the integral in Eq. (\ref{rhos_xi}) is dominated by energies close to the
chemical potential. Hence, we may estimate it by replacing $N\left( \xi
\right) $ and $\left\langle \mathrm{v}_{x}^{2}\right\rangle \left( \xi
\right) $ by their values at the chemical potential (we assume that $\mu $
is not too close to zero in order to avoid the logarithmic divergence of $%
N(\xi )$ at the middle of the band). Changing variables to $\eta =\beta
\sqrt{\xi ^{2}+\tilde{\Delta}^{2}}$, we obtain
\begin{equation}
\rho _{s}\simeq \frac{\tilde{\Delta}^{2}}{T^{2}}N\left( 0\right)
\left\langle \mathrm{v}_{x}^{2}\right\rangle \left( 0\right) \int_{\beta
\tilde{\Delta}}^{\beta W}d\eta \frac{F\left( \eta \right) }{\eta \sqrt{\eta
^{2}-\left( \beta \tilde{\Delta}\right) ^{2}}}\text{,}
\end{equation}%
where $F\left( \eta \right) =\frac{1}{2\eta }\tanh \left( \frac{\eta }{2}%
\right) -\frac{e^{\eta }}{\left( 1+e^{\eta }\right) ^{2}}$. at low $\eta $, $%
F\left( \eta \right) =O\left( \eta ^{2}\right) $, so the integral converges
in the limit $\beta \tilde{\Delta}\rightarrow 0$. At high $\eta $, $F\left(
\eta \right) \sim \frac{1}{\eta }$, so we may also take the $\beta
W\rightarrow \infty $ limit. Therefore, we obtain to leading order in $\beta
\tilde{\Delta}$ and $\frac{1}{\beta W}$
\begin{eqnarray}
\rho _{s}\text{(neg. U sites)} &\simeq &\alpha \frac{\tilde{\Delta}^{2}}{%
T^{2}}N\left( 0\right) \left\langle \mathrm{v}_{x}^{2}\right\rangle \left(
0\right)  \nonumber \\
&\sim &\frac{t_{\perp }^{4}}{\Delta _{0}^{2}T^{2}}N\left( 0\right)
\left\langle \mathrm{v}_{x}^{2}\right\rangle \left( 0\right) ,
\label{rs_low_app}
\end{eqnarray}%
where $\alpha =\int_{0}^{\infty }d\eta F\left( \eta \right) /\eta
^{2}=7\zeta (3)/8\pi ^{2}$, and we have used Eq. (\ref{Dprox}). This is Eq. (%
\ref{rs_low_tperp}). We have verified Eq. (\ref{rs_low_app}) by calculating $%
\rho _{s}$ using finite temperature perturbation theory to order $t_{\perp
}^{4}$, by integrating out the fermions to obtain an effective action for
the superconducting phase, and by evaluating Eq. (\ref{rs}) numerically in
the low $t_{\perp }$ limit.

In the case of superconducting wires, the \textquotedblleft pairing
layer\textquotedblright\ has a finite stiffness of order $\rho _{s}^{x}\sim
t^{\prime }$ 
in the $x$ direction (parallel to the wires) even in the $t_{\perp
}\rightarrow 0$ limit. 
In the transverse direction, however, Eq. (\ref{rs_low_app}) applies for $%
\rho _{s}^{y}$, with the exception that now the proximity induced gap
depends on $\mathbf{k}$.\ The induced gap is significant in a sliver in $%
\mathbf{k}$ space of width $\delta k_{x}\sim \frac{\Delta _{0}}{%
v_{F}^{\prime }}$
around the Fermi surface of the wires (where $v_{F}^{\prime }$ are the Fermi
velocities of the wires), reaching a maximum of order $\frac{t_{\perp }^{2}}{%
\Delta _{0}}$, and negligible elsewhere (since only the region of Fermi
surface of the wires has considerable particle-hole mixing). Taking $%
N(0)\left\langle \mathrm{v}_{y}^{2}\right\rangle \sim t$ and $v_{F}^{\prime
}\sim t^{\prime }\sim t$, we therefore estimate $\rho _{s}^{y}$ in this case
as
\begin{equation}
\rho _{s}^{y}\sim \frac{t_{\perp }^{4}t}{\Delta _{0}^{2}T^{2}}\left( \frac{%
\Delta _{0}}{t^{\prime }}\right) =\frac{t_{\perp }^{4}}{\Delta _{0}T^{2}}.
\end{equation}

The geometric average of $\rho _{s}^{x}$ and $\rho _{s}^{y}$ (which
determines $T_{c}$) scales as
\begin{equation}
\rho _{s}\left( \text{s.c. wires}\right) =\sqrt{\rho _{s}^{x}\rho _{s}^{y}}%
\sim \sqrt{\frac{t}{\Delta _{0}}}\frac{t_{\perp }^{2}}{T},
\label{rs_low_app_wires}
\end{equation}
which is Eq. (\ref{rs_low_wires}).

\subsection{$T_{MF}$ in the low $t_{\perp }$ limit}

$T_{MF}$ is obtained by the equation
\begin{equation}
U\chi _{SC}\left( T_{MF}\right) =1\text{,}  \label{stoner}
\end{equation}%
where $\chi _{SC}\left( T\right) $ is the superconducting susceptibility of
the pairing layer with $\Delta =0$. Using finite temperature perturbation
theory, $\chi _{SC}\left( T_{MF}\right) $ can be expanded in powers of $%
t_{\perp }$. Since we are dealing with a non-interacting theory, all the
diagrams are straight lines with $t_{\perp }$ vertices along them. The
leading order correction to $\chi _{SC}\left( T\right) $ is
\begin{equation}
\delta \chi _{SC}=\frac{2t_{\perp }^{2}a^2}{\beta \Omega}\sum_{i\omega _{n}%
\mathbf{k}}\frac{1}{i\omega _{n}+\varepsilon _{\mathbf{k}}}\frac{1}{\left(
-i\omega _{n}+\varepsilon _{\mathbf{k}}\right) ^{2}}\frac{1}{-i\omega
_{n}+\xi _{\mathbf{k}}},
\end{equation}%
where the dispersions in the pairing and metallic layers are given by $%
\varepsilon _{\mathbf{k}}=-2t^{\prime }\cos k_{x}-(\mu-\varepsilon -\delta
\varepsilon)$, and $\xi _{\mathbf{k}}=-2t\left( \cos k_{x}+\cos k_{y}\right)
-\mu $, respectively. Here, $a^2$ is the unit cell area and $\omega _{n}=%
\frac{\left( 2n+1\right) \pi }{\beta }$ are Matsubara frequencies.
Performing the Matsubara summation, we obtain
\begin{widetext}
\begin{equation}
\delta \chi _{SC}=\frac{2t_{\perp }^{2}a^2}{\Omega}\sum_{\mathbf{k}}\left[ \frac{%
\left[ \varepsilon _{\mathbf{k}}-\xi _{\mathbf{k}}\tanh \left( \beta
\varepsilon _{\mathbf{k}}/2\right) \right] }{4\varepsilon _{\mathbf{k}%
}^{2}\left( \varepsilon _{\mathbf{k}}^{2}-\xi _{\mathbf{k}}^{2}\right) }+%
\frac{2f\left( \varepsilon _{\mathbf{k}}\right) \left[ 1-\beta f\left(
-\varepsilon _{\mathbf{k}}\right) \left( -\varepsilon _{\mathbf{k}}+\xi _{%
\mathbf{k}}\right) \right] }{4\varepsilon _{\mathbf{k}}\left( -\varepsilon _{%
\mathbf{k}}+\xi _{\mathbf{k}}\right) ^{2}}-\frac{f\left( \xi _{\mathbf{k}%
}\right) }{\left( \xi _{\mathbf{k}}+\varepsilon _{\mathbf{k}}\right)
\left( -\xi _{\mathbf{k}}+\varepsilon _{\mathbf{k}}\right)
^{2}}\right]. \label{Qi}
\end{equation}
\end{widetext}

In the case of disconnected negative-$U$ sites, we take the limit $%
\varepsilon _{\mathbf{k}}\rightarrow 0$ in Eq. (\ref{Qi}) (assuming that the
negative-$U$ sites are close to half-filling). The limit gives
\begin{eqnarray}
\delta \chi _{SC}\left( \text{neg. }U\text{ sites}\right) &=&-\frac{%
2t_{\perp }^{2}a^{2}}{\Omega }\sum_{\mathbf{k}}\frac{\beta \xi _{\mathbf{k}%
}-2\tanh \left( \frac{\beta \xi _{\mathbf{k}}}{2}\right) }{4\xi _{\mathbf{k}%
}^{3}}  \nonumber \\
&\approx &-\tilde{\alpha}\frac{t_{\perp }^{2}N\left( 0\right) a^{2}}{T^{2}},
\end{eqnarray}%
where $\tilde{\alpha}=\int_{0}^{\infty }dxx^{-3}\left[ x-2\tanh \left(
x/2\right) \right] =7\zeta (3)/2\pi ^{2}$%
. We have replaced $N\left( \xi \right) \ $by $N\left( 0\right) $, which is
a reasonable approximation since the integral is dominated by the low energy
regime. Adding $\delta \chi _{SC}$ to the zeroth-order susceptibility $\chi
_{SC}^{0}\approx \frac{1}{4T}$ of disconnected sites, we get from Eq. (\ref%
{stoner})
\begin{equation}
\frac{U}{4T_{MF}}\left[ 1-\frac{4\tilde{\alpha}N(0)a^{2}t_{\perp }^{2}}{%
T_{MF}}+O\left( t_{\perp }^{4}\right) \right] =1\text{.}
\end{equation}%
Solving for $T_{MF}$ to leading order in $t_{\perp }^{2}$, we get
\begin{equation}
T_{MF}\text{(neg. }U\text{ sites)}\approx \frac{U}{4}\left[ 1-\frac{%
At_{\perp }^{2}}{tU}+O\left( t_{\perp }^{4}\right) \right] ,
\label{TMF_low_app}
\end{equation}%
where $A=16\tilde{\alpha}tN(0)a^{2}\approx 2\tilde{\alpha}$, where $%
N(a)a^{2}\approx W^{-1}=(8t)^{-1}$ was used. This is Eq. (\ref{TMF_low}).

In the case of superconducting wires, we can still estimate the parametric
form of the most divergent part of $\delta \chi _{SC}$ at low temperatures.
The strongest singularity of the integral in (\ref{Qi}) comes from the
vicinity of the crossing of the two Fermi surfaces (\emph{i.e.} $\xi _{%
\mathbf{k}}=0 $, $\varepsilon _{\mathbf{k}}=0$). This singularity is cut off
by the temperature. As a rough estimation of the integral, we evaluate the
integrand in the limit $\beta \left\vert \varepsilon _{\mathbf{k}%
}\right\vert $, $\beta \left\vert \xi _{\mathbf{k}}\right\vert \gg 1$, so
that $\tanh \left( \frac{ \beta\varepsilon _{\mathbf{k}}}{2}\right)
\rightarrow $ sign$\left( \varepsilon _{\mathbf{k}}\right) $, $f\left(
\varepsilon _{\mathbf{k}}\right) \rightarrow \Theta \left( - \varepsilon _{%
\mathbf{k}}\right) $ where $\Theta $ is a Heaviside step function, etc., and
extend the integration only to within $T$ of the line $\varepsilon _{\mathbf{%
k}}=0$. Further, we change variables from $\mathbf{k}$ to $\left(
\varepsilon _{\mathbf{k}},\xi _{\mathbf{k}}\right) $, with Jacobian $J\left(
\varepsilon ,\xi \right) =\frac{1}{\left\vert \nabla _{\mathbf{k}%
}\varepsilon \times \nabla _{\mathbf{k}}\xi \right\vert }$ which we replace
by its value at $\left( \varepsilon =0,\xi =0\right) $. Adding the
contributions in the four quadrants around the point $\left( \varepsilon
=0,\xi =0\right) $ (both $\varepsilon $ and $\xi $ can be positive or
negative), we get:
\begin{eqnarray}
\delta \chi _{SC} &\sim &-2t_{\perp }^{2}J\left( 0,0\right)
\int_{T}^{W}d\varepsilon \int_{0}^{W}d\xi \frac{1}{2\varepsilon \left(
\varepsilon +\xi \right) ^{2}}  \nonumber \\
&=&2t_{\perp }^{2}J\left( 0,0\right) \left[ \frac{1+\ln \frac{W+T}{2T}}{W}-%
\frac{1}{T}\right]  \nonumber \\
&\sim &-\frac{\tilde{A}t_{\perp }^{2}}{W^{2}T},  \label{dQi2}
\end{eqnarray}%
where we have estimated $J\left( 0,0\right) \sim \frac{1}{W^{2}}$, and kept
only the most divergent term at $T\rightarrow 0$. $\tilde{A}>0$ is a
numerical coefficient. Adding (\ref{dQi2}) to the $t_{\perp }=0$
superconducting susceptibility, which is of the BCS\ form $\chi _{SC}^{0}=%
\frac{1}{2W}\ln \left( \frac{W}{2T}\right) $, we get the following equation
for $T_{MF}$:
\begin{equation}
U\chi _{SC}\left( T_{MF}\right) =U\left[ \frac{1}{2W}\ln \left( \frac{W}{%
2T_{MF}}\right) -\frac{\tilde{A}t_{\perp }^{2}}{W^{2}T_{MF}}\right] =1\text{,%
}
\end{equation}%
and hence we get
\begin{equation}
T_{MF}\text{ (S.C. wires)}\approx T_{MF,0}\left[ 1-\frac{2\tilde{A}t_{\perp
}^{2}}{WT_{MF,0}}+O\left( t_{\perp }^{4}\right) \right],
\label{TMF_app_low_1D}
\end{equation}%
where $T_{MF,0}=\frac{W}{2}e^{-\frac{2W}{U}}$ is the $t_{\perp }=0$ mean
field transition temperature.


\begin{thebibliography}{99}
\bibitem{comment-wires1} This is strictly correct only at zero temperature.
At $t_{\perp }=0$, the wires are completely decoupled, and the phase
stiffness is actually zero for any $T>0$ due to phase fluctuations. However,
for large enough $t_{\perp }$, the mean-field analysis becomes a good
approximation.
\bibitem{comment-wires2} The coupling, $t_\perp$, between the weakly paired
wires and the metal is most effective if a quasiparticle near the Fermi
energy can tunnel from one plane to the other in a momentum conserving
process. For this to be possible, it is necessary that the Fermi surfaces of
the decoupled systems intersect at
crossing points in $\mathbf{k}$ space. If the Fermi surfaces do not cross,
the superfluid density of the composite system is considerably suppressed,
especially for small $t_{\perp}$.
\bibitem{comment-exponents} The fact that $a=1$ can be seen from Eq. (\ref%
{TMF_low}). The derivation of Eq. (\ref{rs_low_Delta}) is similar to that of
$\rho_s$ in the low $t_\perp$ limit (Appendix \ref{sec:app_low_tperp}), and
will be presented elsewhere.
\bibitem{comment-merchant} Strictly speaking, the system never becomes a
full-fledged superconductor in the sense of zero resistance, but there is an
increasingly sharp drop in the resistivity by many orders of magnitude that
begins at the bulk $T_c$ of Pb.
\end{thebibliography}
\end{document}